# Parsimonious Dynamic Mode Decomposition: A Robust and Automated Approach for Optimally Sparse Mode Selection in Complex Systems

Arpan Das[a*], Pier Marzocca[a], Oleg Levinski[b]

[a]*Sir Lawrence Wackett Defence and Aerospace Centre, RMIT University, Melbourne, Victoria, 3082, Australia*
[b]*Defence Science and Technology Group, Fisherman's Bend, Victoria, 3207, Australia*

**Abstract**

This paper introduces the Parsimonious Dynamic Mode Decomposition (parsDMD), a novel algorithm designed to automatically select an optimally sparse subset of dynamic modes for both spatiotemporal and purely temporal data. By incorporating time-delay embedding and leveraging Orthogonal Matching Pursuit (OMP), parsDMD ensures robustness against noise and effectively handles complex, nonlinear dynamics. The algorithm is validated on a diverse range of datasets: purely temporal signals, datasets with hidden dynamics, fluid dynamics simulations (flow past a cylinder and transonic buffet), and atmospheric sea-surface temperature (SST) data. ParsDMD addresses a significant limitation of the traditional sparsity-promoting DMD (spDMD), which requires manual tuning of sparsity parameters through a rigorous trial-and-error process to balance between single-mode and all-mode solutions. In contrast, parsDMD autonomously determines the optimally sparse subset of modes without user intervention, while maintaining minimal computational complexity. Comparative analyses demonstrate that parsDMD consistently outperforms spDMD by providing more accurate mode identification and effective reconstruction in noisy environments. These advantages render parsDMD an effective tool for real-time diagnostics, forecasting, and reduced-order model construction across various disciplines.

**Nomenclature**

Corresponding author
*Arpan Das, Research Assistant, RMIT University, Melbourne, Victoria 3082, Australia
Email: arpand1989@gmail.com

| | | |
|---|---|---|
| $A$ | = | Linear operator |
| $\tilde{A}$ | = | Projection of $A$ onto POD modes |
| $B$ | = | Basis Matrix |
| $b$ | = | Vector coefficients of DMD modes |
| $b_k$ | = | Initial amplitude of DMD modes |
| $b_{opt}$ | = | Optimal mode amplitudes |
| $\beta$ | = | Aspect ratio of snapshots |
| $C$ | = | Measurement matrix |
| $D_b$ | = | Diagonal matrix containing amplitude of modes |
| $\gamma$ | = | Sparsity value |
| $J(b)$ | = | Cost function of $b$ |
| $\Lambda$ | = | Eigen values matrix |
| $\lambda_0$ | = | Hyperparameter for Augmented Lagrange Multiplier |
| $m$ | = | Number of snapshots taken |
| $n$ | = | Spatial points per snapshot |
| $\Omega$ | = | Diagonal matrix of eigenvalues |
| $\eta$ | = | Noise magnitude |
| $\Phi$ | = | DMD Modes |
| $\tilde{\Phi}$ | = | Reduced DMD modes |
| $r$ | = | Rank of SVD |
| $\Sigma$ | = | Matrix containing singular values |
| $\tau$ | = | Optimal threshold |
| $U$ | = | Left Singular Vector |
| $\mu_\beta$ | = | Median Marcenko-Pastur distribution |
| $V$ | = | Right Singular Vector |
| $V_\lambda$ | = | Vandermonde matrix |
| $W$ | = | Eigen vector |
| $X$ | = | Data matrix |
| $\tilde{X}$ | = | POD projected data matrix |
| **DMD** | = | Dynamic Mode Decomposition |
| **POD** | = | Proper Orthogonal Decomposition |
| **RPCA** | = | Robust Principal Component Analysis |
| **SVD** | = | Singular Value Decomposition |

# 1. Introduction

The field of data-driven modeling and control in complex systems is progressing at a rapid pace, offering the immense potential to revolutionize the engineering, biological, and physical sciences. The availability of high-fidelity measurements from historical records, numerical simulations, and experimental data is unprecedented. However, despite the abundance of data, constructing accurate models continues to be a challenging endeavor. Contemporary systems of interest, including turbulent flows, epidemiological systems, neuroscience, financial markets, or the climate, can be characterized as high-dimensional, nonlinear dynamic systems that demonstrate diverse multiscale spatiotemporal phenomena. Despite their complexity, numerous systems exhibit evolution on a low-dimensional attractor which can be described by spatiotemporal coherent structures.

Dynamic Mode Decomposition (DMD) was originally developed in the fluid dynamics community [1] as a method to analyze complex flows [2] by decomposing the data into a set of dynamic modes based on spatiotemporal coherent structures. Dynamic Mode Decomposition (DMD) is a recent extension of the classical Arnoldi technique [3–6] to accommodate a data-based rather than a model-based framework [1,2]. A Krylov sequence of flow fields is subjected to a high-degree polynomial fitting, assuming that the flow fields will eventually become linearly dependent after a certain number of snapshots. Once this condition is met, it is assumed that there exists a general linear dependence among the snapshots, and the mapping between snapshots (which is unknown) is expressed within the basis formed by the snapshots. The optimal linear combination of these snapshots represents a low-dimensional representation of the system dynamics, which allows for subsequent analysis, such as stability and receptivity computations, to be applied to it.

The broad applicability and the potential of DMD inspired the research community to contribute to the development of a more robust DMD algorithm for various strategic uses. Tu *et al.* [7] presented a theoretical framework that defined DMD as the eigen decomposition of an approximate linear operator. The primary difference between this [7] framework and the original [1] is how the DMD modes are formulated. Tu *et al.* [7] formulated the DMD modes as the exact eigenvectors of the linear operator, therefore called as Exact DMD, whereas Schmid [1], it was formulated as projected DMD modes.

DMD is known for its wide applicability due to its connection with Koopman spectral theory [8–10]. The Koopman operator, an infinite-dimensional linear operator, is utilized to describe the evolution of measurements in non-linear dynamical systems. As these measurements are in the form of functions, they establish an infinite-dimensional Hilbert space, thereby computationally intractable. Rowley et al. [11] have illustrated that under certain conditions, DMD can present a finite-dimensional approximation of the infinite-dimensional Koopman operator, proving to be highly beneficial for the analysis of intricate nonlinear systems. Although, the selection of observables is a critical factor in this process [12].

Since its development in the fluid dynamics community, DMD has been utilized in analyzing several complex system dynamics across a wide variety of engineering fields [13–23], atmospheric sciences [24–27], physics [28–33], biology [34,35], neuroscience [36,37], epidemiology [38,39], robotics [17], financial trading strategies [40] among others. Established over a decade ago, DMD, like any other approach, has experienced several modifications to improve its resilience in managing noise, nonlinearity, nonstationarity, among other challenges [13,26,27,41–47].

One key factor in the analysis established by the DMD method is to identify the dominant dynamic modes, and to achieve a trade-off between accuracy and complexity of the model. Sparsity-promoting DMD (spDMD) [48] introduced a method which penalizes the $\ell_1$-norm of the vector of DMD mode amplitudes. Although it is an effective method in identifying a sparse subset of DMD modes, it has some limitations. Due to the nature of the spDMD algorithm which utilizes the ADMM algorithm, it requires manual identification of the sparsity values that leads to a 1-mode solution and $r$-mode solution (provided '$r$' is the SVD rank). Moreover, these values vary drastically between different dynamical systems, even with the same system contaminated with noise. Furthermore, spDMD struggles to identify the true rank of the system when data is contaminated with noise by overestimating the number of DMD modes.

In this paper, a novel DMD method named Parsimonious Dynamic Mode Decomposition is showcased. It has an in-built implementation of time-delay embedding capability on the POD projected data, thereby rendering it a fast and robust method against noise and nonlinearity. The parsimony of this technique is achieved through the automatic identification of an optimally sparse subset of DMD modes. This is facilitated by employing principles from Orthogonal Matching Pursuit (OMP), a greedy algorithm that iteratively solves a system of equations, selecting the coefficient that maximizes the projection with the current residual at each step. Additionally, the algorithm's convergence is fine-tuned through a custom scaling parameter that balances sparsity and accuracy. This method is demonstrated across various dynamical systems: synthetic time-series data with high amount of noise, spatiotemporal data with hidden dynamics buried under noise, fluid flow problems (classical low-Re flow past cylinder and significantly complicated transonic shock buffet), and also its application in atmospheric science by analyzing sea-surface temperature (SST) data. A comparative analysis with sparsity-promoting DMD is also conducted, highlighting the advantages of the parsDMD algorithm. The findings indicate that this method excels rapidly and automatically identifying sparse mode sets, thereby serving as a valuable tool for diagnostics, forecasting, and the development of reduced-order models.

## 2. Exact DMD Algorithm and Mode Selection Criteria

The DMD framework operates on the principle of perceiving a complex dynamical system without relying on explicit equations. Instead, it utilizes measurements obtained from experimental tests or numerical simulations to approximate the system's dynamics and forecast future states. This is achieved by constructing a proxy, which represents an approximate locally linear dynamical system. Although first discovered by Schmid [1,49], a more modern definition is given by Tu et al. [7].

*2.1 Exact DMD algorithm*

The definition states that considering a dynamical system and two sets of data:

$$X = [x_1\ x_2\ x_3\ ...\ x_{m-1}] \tag{1}$$

$$X' = [x'_1\ x'_2\ x'_3\ ...\ x'_{m-1}] \tag{2}$$

such that $x'_k = F(x_k)$, where the map $F$ represents the evolution of the dynamical system in time $\Delta t$, and each snapshot $x_k$ contains $n$ measurements, and $m$ snapshots. DMD computes the leading eigen-decomposition of the best-fit linear operator $A$ relating the data $X' = AX$. Consequently, the matrix $A$ can be computed by obtaining the pseudo-inverse of $X$ and then performing a matrix multiplication with $X'$, such that,

$$A = X'X^\dagger \tag{3}$$

Typically, the state dimension n is often large, leading to difficulties in computing the matrix $A$. To address this issue, a rank-reduced representation involving a POD-projected matrix $\widetilde{A}$ is computed as an alternative to directly performing the eigen-decomposition of $A$. As a result, the procedure for conducting Dynamic Mode Decomposition (DMD) begins with the Singular Value Decomposition (SVD) of $X$, such that,

$$X \approx U\Sigma V^* \tag{4}$$

where * represents conjugate transpose, $U \in \mathbb{C}^{n \times r}, \Sigma \in \mathbb{C}^{r \times r}$, and $V \in \mathbb{C}^{m \times r}$ where $r$ is the rank of the reduced SVD approximation of $X$. The SVD reduction at the first stage is performed to enable a low-rank truncation of the data. Although matrix $A$ can be directly computed using the pseudoinverse of $X$ obtained from the SVD, it is more efficient to compute the reduced linear operator $\widetilde{A}$ onto the POD modes:

$$\widetilde{A} = U^* A U = U^* X' V \Sigma^{-1} \tag{5}$$

Consequently, the eigen-decomposition of $\widetilde{A}$ is computed:

$$\widetilde{A} W = W \Lambda \tag{6}$$

The eigenvectors are represented by the columns of matrix $W$, while the corresponding eigenvalues $\lambda_k$ are located along the diagonals of matrix $\Lambda$. The final stage involves reconstructing the eigen-decomposition of matrix of $A$ from $W$ and $\Lambda$ is performed. The eigenvalues are given by $\Lambda$ and the eigenvectors (the DMD modes) are given by the columns of $\Phi$:

$$\Phi = X' V \Sigma^{-1} W \tag{7}$$

Consequently, by utilizing the low-rank approximation of eigenvalues and the eigenvectors, and introducing $\omega_k = \ln(\lambda_k)/\Delta t$ for continuous eigenvalues, the projected solution for all future temporal states can be found as,

$$x(t) \approx \sum_{k=1}^{r} \phi_k \exp(\omega_k t) b_k = \Phi \exp(\Omega t) \mathbf{b} \tag{8}$$

where $b_k$ is the initial amplitude of each mode, $\Phi$ is the matrix having columns as the DMD eignevectors $\phi_k$, and $\Omega = diag(\omega)$ is a diagonal matrix in which the diagonal entries are the eigenvalues $\omega_k$. The eigenvectors $\phi_k$ shares the same size as the state $x$, and $\mathbf{b}$ is a vector containing the coefficients $b_k$.

2.2 *Mode Selection Criteria*

Several methods can be applied to select the leading DMD modes. The modes can be selected based on their amplitudes. Another way of selecting modes can be based on energy thresholding, where the modes are selected based on their contribution to the total energy of the system by computing their energy content and sorting them in descending order. Consequently, a hyperparameter or a threshold criterion can be utilized to select a subset of modes among all the modes. Kou and Zhang [50] computed the time coefficient of the modes, which at a particular instant represents the mode's energy at that instant. Integration of the time coefficient over the entire sample of the dataset represents the energy contribution of the modes towards the system dynamics. Afterwards, a hyperparameter can be used to filter out the few dominant modes among the entire set of sorted modes according to their energy contribution. Moreover, if the system exhibits distinct frequency bands, DMD modes can be filtered based on their frequency content. This can be done using frequency domain techniques. Sometimes, the selection of modes is guided by physical insight into the system dynamics. Moreover, sparsity-promoting techniques can be applied to select a subset of modes [48].

In order to compute $x(t)$ described in Eq. (8), it is imperative to determine the coefficient values $b_k$. The most straightforward way to achieve that is to consider the initial snapshot $x_1$ at time $t_1 = 0$, so that $x_1 = \Phi b$. In general, the matrix of eigenvectors $\Phi$ is not a square matrix, therefore, b can be computed as,

$$\mathbf{b} = \Phi^\dagger x_1 \tag{9}$$

where $\Phi^\dagger$ is the pseudoinverse of $\Phi$. Consequently, the modes can be ordered according to their respective mode amplitudes. Despite its simplicity, this method calculates the mode amplitudes solely using the initial snapshot. In the event of a nonlinear or noise contaminated dataset, it is highly likely that $x_1$ may not reside within the column space of $X'$. This occurrence is quite common in practical scenarios. Therefore, an alternate formulation is utilized [48] to construct a cost function the Frobenius norm between the original and the reconstructed dataset, such that,

$$\underset{b}{\text{minimize}}\, J(\mathbf{b}) = \|X - \Phi D_b V_\lambda\|_F^2 \tag{10}$$

where $D_b$ is a diagonal matrix containing the mode amplitudes, and $V_\lambda$ is a Vandermonde matrix with eigenvalues. Solving this optimization problem the mode amplitudes can be identified. However, it should be mentioned that due to the least-square nature of the computations, all of the modes will have non-zero amplitudes, and can be further truncated using some hyperparameter to threshold the selection of modes. Jovanovic *et al.* [48] incorporated sparsity on Eq. (10) to identify a sparse subset of modes, such that,

$$\underset{b}{\text{minimize}}\, J(\mathbf{b}) = \|X - \Phi D_b V_\lambda\|_F^2 + \gamma \|b\|_1 \tag{11}$$

where $\| \, . \, \|_1$ denotes the $\ell_1$ norm penalization that promotes sparsity. Consequently, (Eq. 11) is solved using Alternating Direction Method of Multipliers (ADMM) algorithm. Although, it should be noted that to identify the optimally sparse set of modes, the sparsity parameter $\gamma$ pertaining to 1-mode solution and $r$-mode (based on SVD rank truncation $r$) needs to be identified. Consequently, the algorithm computes the mode amplitudes for a set of sparsity values that lies in between the extreme values set by the user, and the optimal mode amplitudes are calculated. For more details, the readers are directed to [48]. Although, this algorithm provides a sparse subset of modes, identifying the sparsity two extreme sparsity parameters is a hit and trial process and varies wildly among different datasets [19,48].

Another formulation of the cost function was developed [27] by Askham and Kutz that utilizes nonlinear least squares to compute the DMD and significantly reduce the bias due to noise:

$$\underset{\omega,\phi_b}{\text{minimize}}\, J(\mathbf{b}) = \|X - \mathbf{\Phi_b}\mathbf{T}(\omega)\|_F^2 \tag{12}$$

DMD mode selection procedure for the abovementioned methods and several other DMD methods [41–43] are not automatic and require human intervention to identify the DMD modes among a set of $r$ modes. In the following section, a parsimonious DMD methodology is proposed that can be used for totally automatic and optimally sparse mode selection.

## 3. Parsimonious DMD

The DMD method enables the decomposition of data into dynamic modes, which capture both spatial and temporal characteristics. It was first discovered by Schmid [51,52] and is generally referred to as Standard DMD. A more modern definition called Exact DMD is given by Tu et al. [7]. The two algorithms are either practically the same or similar except for the DMD mode formulation, although they provide almost identical results. For the sake of conciseness, while addressing these methods, they are referred to as Standard DMD in this paper. Although, a powerful method for spatiotemporal decomposition of data, standard DMD is unable to capture purely temporal data [41,53], is highly sensitive to noise [44], and doesn't perform well in addressing nonlinear dynamics of a complex system [41]. To address these issues and to use the parsimonious DMD method on datasets belonging to a broader class of dynamical systems, time-delay embedding is incorporated into the Parsimonious DMD methodology showcased in this paper. Therefore, Takens' delay embedding theorem [54] is incorporated with the DMD algorithm. The expansion resulting from time-delay embedding is suitable for representing different types of nonlinear dynamical systems, including attractors, dynamics deviating from attractors due to instabilities, and transient decay caused by attractors [41]. The entire procedure for performing Parsimonious Dynamic Mode Decomposition can be outlined as follows:

*3.1. SVD and Truncation*

At first, a Singular Value Decomposition (SVD) of the entire data matrix **Y** [53,55] is computed. This is necessary because if the time-delay embedding of the data matrix is performed first, it would be computationally challenging to compute the SVD of the delay-embedded matrix in most cases, particularly when dealing with moderately high-dimensional measurements. Hence, the SVD is carried out on the entire set of measurements **Y**, such that,

$$\mathbf{Y} \approx \mathbf{U\Sigma V}^* \tag{13}$$

After computing SVD of the data, it is necessary to perform a rank truncation based on the computed singular values. The determination of this choice is influenced by multiple factors, including the source, quality of data, and the dynamic significance of low-energy modes. Generally, due to the complexity associated with different dynamical systems, it is quite often that the truncation rank for SVD is chosen differently. Although there exist a few avenues that can be utilized to approximate the rank truncation automatically.

Gavish and Donoho [56] have provided a theoretical framework for achieving optimal singular value truncation, even when there is additive white noise error, among the different threshold criteria. The threshold criterion is as follows,

$$\tau = \lambda(\beta)\sqrt{n}\eta \tag{14}$$

where $\beta = n/m$ is the aspect ratio of the data matrix and $\lambda(\beta)$ is,

$$\lambda(\beta) = \left(2(\beta + 1) + \frac{8\beta}{(\beta+1)+(\beta^2+14\beta+1)}\right)^{1/2} \tag{15}$$

Generally, the noise magnitude $\eta$ is unknown and estimated directly from the SVD of the data matrix. In that case, the optimal threshold $\tau$ is given by,

$$\tau = \omega(\beta)\sigma_{median} \tag{16}$$

where $\omega(\beta) = \lambda(\beta)/\mu_\beta$ and $\mu_\beta$ is the median of the Marcenko-Pastur distribution [56], respectively.

Another widely used criterion for singular value thresholding is to use the energy criterion, let's say, the data matrix is of size $mxn$, where $m > n$, then an energy threshold criterion $\epsilon$ can be set such that the cumulative energy of the first $r$ modes contain that amount of energy among the total number of $n$ modes, such that,

$$\frac{\sum_{i=1}^{r} \sigma_i^2}{\sum_{j=1}^{n} \sigma_j^2} \approx \epsilon \tag{17}$$

where $\sigma$ are the singular values.

*3.2 Projection and Delay-embedding*

After the rank truncation, the data is projected onto its rank truncated Proper Orthogonal Decomposition (POD) modes, such that,

$$\widetilde{X} = U^*Y = \Sigma V^* \tag{18}$$

As already stated, the concept of time-delay embedding integrates the principles of Standard DMD with Takens' delay embedding theorem [54]. This integration allows for the capture of phase information related to a purely temporal signal's eigenvalue pair. Additionally, time-delay embedding aids in enhancing the rank of the data matrix when the state measurement is of low dimensionality. This technique can be applied to both over-determined and under-determined data matrices. Consequently, the application of time-delay embedding to the reduced matrix $\widetilde{X}$ can be represented as,

$$\widetilde{X}_{aug} = \begin{bmatrix} \widetilde{x}_1 & \widetilde{x}_2 & \cdots & \widetilde{x}_{m-d+1} \\ \widetilde{x}_2 & \widetilde{x}_3 & \cdots & \widetilde{x}_{m-d} \\ \vdots & \vdots & \ddots & \vdots \\ \widetilde{x}_d & \widetilde{x}_{d+1} & \cdots & \widetilde{x}_m \end{bmatrix} \tag{19}$$

where $\widetilde{X}_{aug}$ is the reduced-order time-delay augmented matrix with delay-order $d$. It should be noted that for $d = 1$ the algorithm reduces to standard DMD. The time-delayed reduced data-matrix $\widetilde{X}_{aug}$ is divided into two sets $\widetilde{X}_1$ and $\widetilde{X}_2$, similar to standard DMD algorithm such that each contains $m - d$ number of snapshots,

$$\widetilde{X}_2 = \widetilde{A}\widetilde{X}_1 \tag{20}$$

where $\widetilde{A}$ is the reduced linear operator.

*3.3 Reduced Linear Operator*

As the data matrix is already rank truncated in the beginning, the reduced linear operator can be computed directly:

$$\widetilde{A} = \widetilde{X}_2 \widetilde{X}_1^{\dagger} \tag{21}$$

Although, depending on the delay-order the size of the augmented matrix might still be large enough for efficient computation of $\widetilde{A}$ directly. Therefore, another SVD on the reduced data-matrix $\widetilde{X}_1$ can be calculated, such as,

$$\widetilde{X}_1 = U_1 \Sigma_1 V_1^* \tag{22}$$

where $U_1, \Sigma_1, V_1$ are rank truncated left singular vector, diagonal matrix containing the singular values, and right singular vectors respectively based on some suitable rank truncation criteria (for example, energy thresholding). Consequently, and the reduced linear operator $\widetilde{A}$ is computed as,

$$\widetilde{A} = U_1^* X_2 V_1 \Sigma_1^{-1} \tag{23}$$

*3.4 Eigendecomposition and DMD Modes*

The eigenvalue decomposition of $\widetilde{A}$ can be computed directly as,

$$\widetilde{A} W = W \Lambda \tag{24}$$

Although, in this work, the abovementioned direct eigendecomposition of $\widetilde{A}$ is used if time-delay embedding is incorporated ($d > 1$), otherwise the eigenvalues and eigenvectors are computed based on the Total Least Squares criterion [57]. After the eigendecomposition of $A$ shown in Eq. (24), the eigenvalues are given by $\Lambda$ and the eigenvectors (reduced DMD modes) are represented by the columns of $\widetilde{\Phi}$:

$$\widetilde{\Phi} = \widetilde{X}_2 V_1 \Sigma_1^{-1} W \tag{25}$$

In case time-delay embedding is not used ($d = 1$), and if the data is noise contaminated there might be a deviation of the identified $\widetilde{A}$ matrix, which might lead to inaccuracies in eigenvalues and eigenvectors computation [44]. In that case, the Total Least Squares algorithm [57] is used to identify the eigenvalues and eigenvectors. The implementation of this method is quite straightforward. At first, the data matrices $\widetilde{X}_1$ and $\widetilde{X}_2$ are stacked on top of each other and as the SVD of the stacked matrix is computed:

$$Z = \begin{bmatrix} \widetilde{X}_1 \\ \widetilde{X}_2 \end{bmatrix} = U_Z \Sigma_Z V_Z^* \tag{26}$$

The left singular vector is divided into four quadrants, such as,

$$U_Z = \begin{bmatrix} U_{11} & U_{12} \\ U_{21} & U_{22} \end{bmatrix} \tag{27}$$

Subsequently, the reduced linear operator is computed as,

$$A_Z = U_{21}U_{11}^\dagger \tag{28}$$

The eigenvalues and DMD modes can be computed directly from the $A_Z$:

$$A_Z\widetilde{\Phi} = \widetilde{\Phi}\Lambda \tag{29}$$

In the final step, the reduced DMD mode is projected back to full-order DMD mode:

$$\Phi = U_1 \widetilde{\Phi} \tag{30}$$

*3.5 Parsimonious Mode Selection*

The standard/exact DMD method involves the computation of the mode amplitude vector **b** by obtaining the pseudoinverse of $\Phi$ and subsequently multiplying it with the first snapshot $x_1$ as shown in Eq. (9). This approach is considered as the most direct way for computing **b**. Nevertheless, it is essential to acknowledge that $x_1$ may not reside within the column space of $X'$, particularly in the presence of nonlinearity in the data. Hence, the identification of an optimal **b** assumes great significance, which is achieved through the minimization of a cost function $J(\mathbf{b})$,

$$\underset{b}{\text{minimize}}\, J(\mathbf{b}) = \|X - \Phi D_b V_\lambda\|_F^2 \tag{31}$$

where $D_b$ is a diagonal matrix that encompasses the amplitudes of the modes in its diagonal entries, and $V_\lambda$ represents a Vandermonde matrix that possesses eigenvalues. The product $\Phi D_b V_\lambda$ is an alternative representation for $x(t)$ as in Eq. (8). By leveraging matrix trace properties, the function $J(\mathbf{b})$ can be expressed in a different form, such as,

$$J(\mathbf{b}) = b^*Pb - q^*b - b^*q + s \tag{32}$$

where, $P = (\Phi^*\Phi) \circ (\overline{V_\lambda V_\lambda^*})$, $q = \overline{diag\{V_\lambda X^*\Phi\}}$, and $s = trace(\Sigma^*\Sigma)$. The objective function $J(\mathbf{b})$, in our case, is somewhat different from [58] as the definition of exact DMD is utilized to compute the reduced order DMD modes (Eq. 12), which in turn was used in deriving the cost function, whereas in the paper by Jovanovic et al. [48] projected DMD $\Phi = UW$ was used. The objective function is minimized at the point where its gradient vanishes. Consequently, the optimal least-square solution can be acquired analytically as,

$$\mathbf{b}_{optimal} = P^{-1}q \tag{33}$$

The solution from (Eq. 33) uses least-square computations, therefore, all the modal amplitudes have a certain weightage associated with them. Recall, that the approach taken by [48] requires identifying the extremum sparsity parameters for 1-mode and $r$-mode solution, and the value of sparsity parameter $\gamma$ pertaining to these extremum values varies wildly from one case to another and the only way to identify them is through using the hit and trial approach. So, it is not possible to automatically compute the sparse subset of mode amplitudes using the method described in [48].

In order to identify the sparsely optimal modes, the primary objective is to:

$$minimize \: \|b\|_0 \: such \: that \: Pb = q \tag{34}$$

As the formulation for (Eq. 34) is non-convex and NP-hard, there is no direct solution. Therefore, to achieve automatic identification of the optimally sparse subset of modes, the idea from Orthogonal Matching Pursuit (OMP) algorithm [59] is utilized. OMP, primarily used for compressed sensing [59] solves a system of equations iteratively, where each iteration identifies the best possible coefficient in that iteration, and number of coefficients identified is equal to the number of iterations solved using OMP algorithm. The detailed procedure for identifying the modes is as follows:

<u>1<sup>st</sup> iteration ($i = 1$)</u>: Let us denote $\boldsymbol{P} = [\widetilde{\boldsymbol{P}}_1 \: \widetilde{\boldsymbol{P}}_2 \: ... \: \widetilde{\boldsymbol{P}}_r]$, and $\boldsymbol{b} = [b_1 \: b_2 \: ... \: b_r]$, where $\widetilde{\boldsymbol{P}}_j$ are the columns of matrix $\boldsymbol{P}$, and $b_j$ are the mode amplitudes contained in column vector $\boldsymbol{b}$. Then we can find the maximum projection of the column $\widetilde{\boldsymbol{P}}_j$ on $\boldsymbol{q}$:

$$argmax_j \: |\widetilde{\boldsymbol{P}}_j^T \boldsymbol{q}| \tag{35}$$

and the identified column is used to construct the basis **B** at iteration 1, such that,

$$\mathbf{B}_{(1)} = [\widetilde{\boldsymbol{P}}_{i(1)}] \tag{36}$$

where $\widetilde{\boldsymbol{P}}_{i(1)}$ denotes the $jth$ column of $\boldsymbol{P}$ that has maximum projection on $\boldsymbol{b}$. So that it minimizes $\|\boldsymbol{q} - \mathbf{B}_{(1)} \boldsymbol{b}^{(1)}\|_2^2$. Therefore, $\boldsymbol{b}^{(1)}$ can be identified as:

$$\boldsymbol{b}^{(1)} = \left(\mathbf{B}_{(1)}^T \mathbf{B}_{(1)}\right)^{-1} \mathbf{B}_{(1)}^T \boldsymbol{q} \tag{37}$$

and the residual can be calculated as,

$$\boldsymbol{res}(1) = \boldsymbol{q} - \mathbf{B}_{(1)} \boldsymbol{b}^{(1)} \tag{38}$$

After that, the 2-norm of the residual is computed:

$$resnorm(1) = \|\boldsymbol{q} - \mathbf{B}_{(1)} \boldsymbol{b}^{(1)}\|_2 \tag{39}$$

The first iteration identifies 1 mode, therefore it would have the maximum error for the cost function $J(b)$, as more iterations are performed, the number of modes will increase, and consequently the reconstruction error will fall. Our objective is to find the optimally sparse subset of modes, i.e. the parsimonious model which will have a trade-off between accuracy and number of modes. In this case, the sparsity parameter is denoted by the number of iterations, therefore, the 1$^{st}$ iteration is the most sparse, and the $r$th iteration pertains to the least sparse (equivalent to the least-square solution) model. In order to compare the reconstruction error with the sparsity, both need to be scaled accordingly. The scaled residual at '$i$th' iteration is:

$$res_{scaled}(i) = resnorm(i)/resnorm(1) \tag{40}$$

such that $0 < res_{scaled} \leq 1$. Although, the most straightforward way to estimate the sparsity value is to scale it linearly $(i/r)$, but it's not quite robust. Especially, because the linear scaling is quite sensitive to SVD rank truncation value $r$ (showcased in Section 4). The primary reason for this phenomenon is that the reconstruction error drops drastically with the first few dominant DMD modes, and then it drops at a much slower rate and almost plateaus. Therefore, the scaled sparsity parameter is formulated as:

$$\zeta(i) = \frac{\log(i+1)}{\log(r+1)} \tag{41}$$

Therefore, a lower value of $\zeta$ indicates higher sparsity, and a higher value of $\zeta$ indicates lesser sparsity. Finally, the distance between the two curves representing scaled residual and sparsity is computed:

$$\delta(i) = \|res_{scaled}(i) - \zeta(i)\|_2 \tag{42}$$

<u>2$^{nd}$ iteration ($i = 2$)</u>: At first, similar to the first iteration, the projection of the columns of $\boldsymbol{P}$ is computed, but instead of identifying the maximum projection on $\boldsymbol{q}$, the maximum projection on the residual is identified, such that,

$$argmax_j\ |\widetilde{\boldsymbol{P}}_j^T res(\boldsymbol{1})| \tag{43}$$

and the basis matrix is updated:

$$\mathbf{B}_{(2)} = [\widetilde{\boldsymbol{P}}_{i(1)}\ \widetilde{\boldsymbol{P}}_{i(2)}] \tag{44}$$

Consequently, the first two mode amplitudes are calculated, such that,

$$\boldsymbol{b}^{(2)} = \left(\mathbf{B}_{(2)}^T \mathbf{B}_{(2)}\right)^{-1} \mathbf{B}_{(2)}^T \boldsymbol{q} \tag{45}$$

where $b^{(2)}$ contains the best two mode amplitudes among the set of $r$ modes. Subsequently, the residual, norm of residual, scaled residual, scaled sparsity, and the distance is computed similarly as shown in Eq. (38)-(42).

<u>$k$th iteration $(i = k)$</u>: The maximum projection of columns of $P$ on the residual from the previous iteration is computed:

$$argmax_j \ |\widetilde{P}_j^T res(k-1)| \tag{46}$$

Consequently, the basis matrix is updated:

$$\mathbf{B}_{(2)} = [\widetilde{P}_{i(1)} \ \widetilde{P}_{i(2)} \ ... \ \widetilde{P}_{i(k)}] \tag{47}$$

and the mode first $k$ mode amplitudes are computed:

$$b^{(k)} = \left(\mathbf{B}_{(k)}^T \mathbf{B}_{(k)}\right)^{-1} \mathbf{B}_{(k)}^T q \tag{48}$$

After that, the residual is computed:

$$res(k) = q - \mathbf{B}_{(k)} b^{(k)} \tag{49}$$

Subsequently, the residual norm, scaled residual, and the corresponding distance between the two curves are computed. It is obvious that as the error decreases and scaled sparsity $\zeta$ increases, the distance between the two curves will converge till it reaches the sparsity value representing the most parsimonious model, and then it will start to diverge. Therefore, a simple stopping criterion is placed to identify the optimally sparse solution, such that,

$$if \ \delta(k) > \delta(k-1) \tag{50}$$

the algorithm stops, $b^{(k)}$ will contain the amplitude of the best possible $k$ modes and the rest of the $r - k$ modes will have zero amplitude.

The entire algorithm is described in the following flowchart illustrated in Fig. 1.

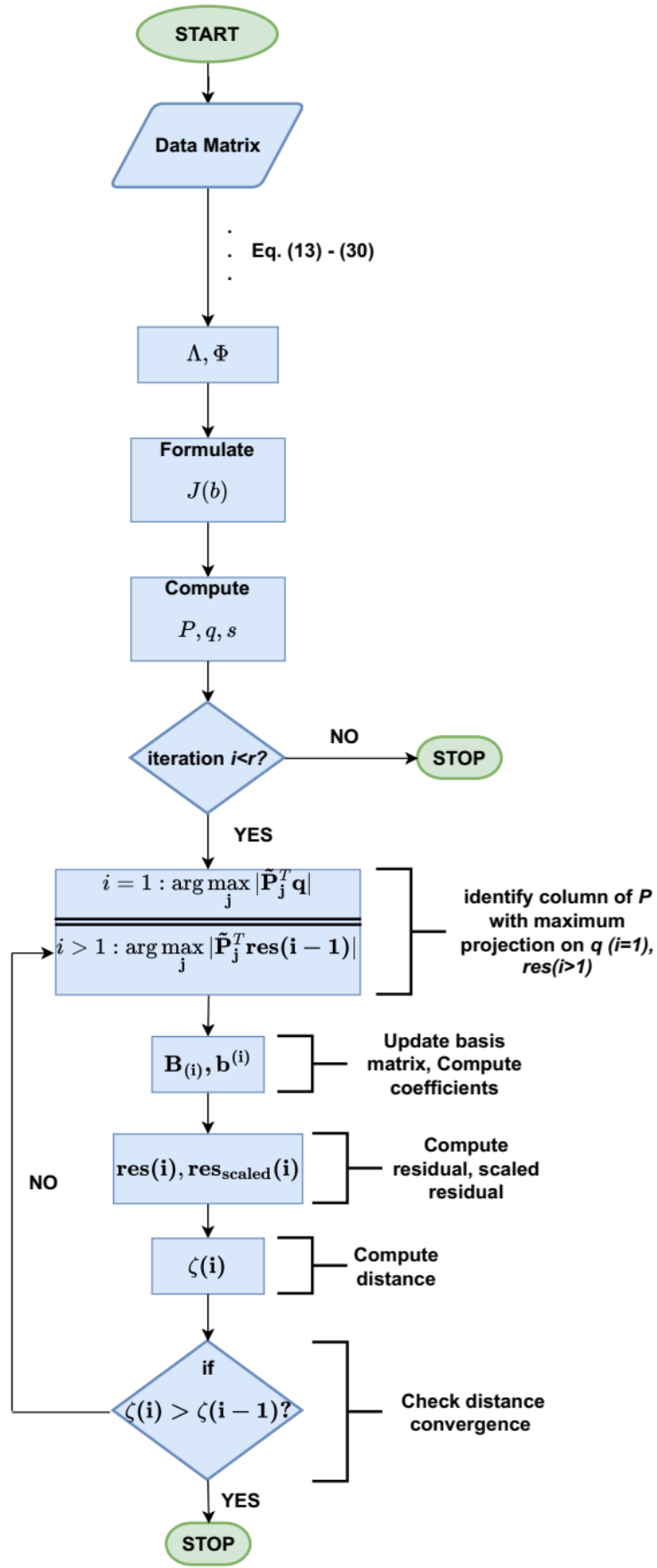

**Fig. 1**: Parsimonious DMD algorithm

## 4. Results and Discussion

A wide range of test cases are showcased to demonstrate the efficacy of the developed algorithm.

*4.1 Synthetic Test Cases*

Toy examples of time-series and spatiotemporal dataset are constructed to showcase the effectiveness of the Parsimonious DMD algorithm.

*4.1.1 Synthetic time series*

A signal is constructed and added Gaussian noise with zero mean ($\mu$) and different levels of variances ($\sigma^2$), such that,

$$f_{clean} = 5\sin(7*2\pi t) + 9\sin(9*2\pi t) + 11\sin(13*2\pi t) \tag{51}$$

$$f_{noisy} = f_{clean} + \mathcal{N}(\mu, \sigma^2) \tag{52}$$

The clean signal and its noisy counterparts are shown in Fig. 2a and 2b respectively. The noisy signals are shifted on the plots for better visualization.

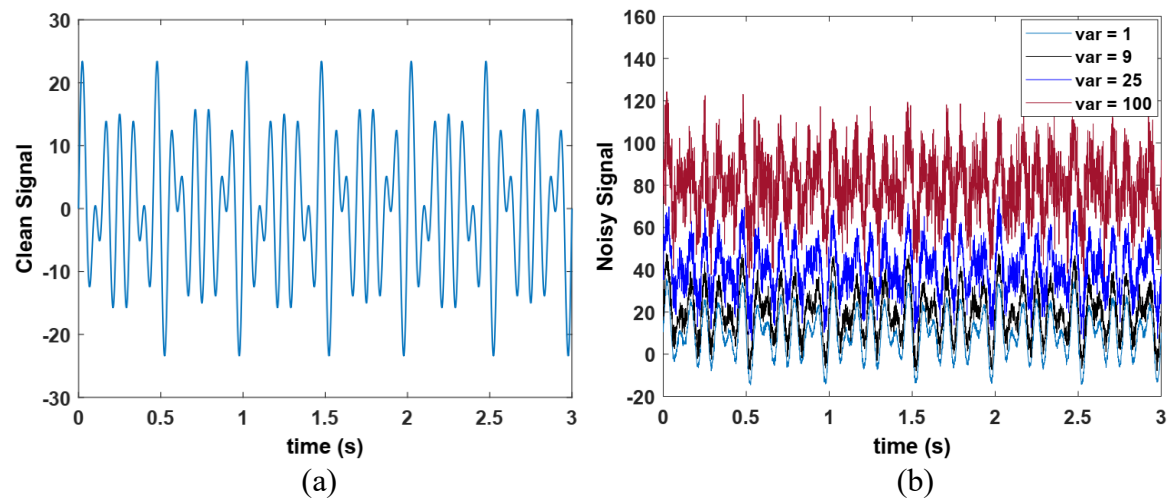

**Fig. 2**: (a) clean signal (b) noisy signal with different variance values

It should be noted that neither the standard nor the exact DMD method [1,53] are capable of handling purely temporal data, therefore, time-delay embedding is utilized to create a Hankel matrix [41,53] from the synthetic signal and consequently DMD is implemented. However, depending on the complexity of the signal (number of frequencies associated, nonlinearity, noise corruption), it is very likely that the time-delay order '$d$' is much larger than the actual number of modes associated with the signal. For example, the synthetically generated signal $f_{clean}$ denoted by Eq. (51) has three frequencies 7, 9, and 13 Hz. Therefore, ideally, a time-delay order of six (since DMD modes are complex conjugate of each other) would identify this signal but instead a delay order $d = 20$ is required to faithfully capture these frequencies. Moreover, the noise counterpart $f_{noisy}$ with high amount of noise would require an astonishingly high delay order of $d = 600$ to capture these frequencies for a faithful reconstruction and future-state prediction. Therefore, although the actual structure of the dynamics is sparse, these six original modes are buried among a large number of spurious modes. The detection of these sparse modes is essential for the effective reconstruction of the original signal and for forecasting, representing a critical task in the development of reduced-order models. A comparison of parsimonious DMD with time-delay embedded Exact DMD is illustrated in Fig. 3 with the help of the DMD spectrum. It is clearly evident that in Exact DMD, the actual modes are buried within hundreds of spurious modes. Even if the modes are ranked according to mode amplitudes based on Eq. (9) or Eq. (10), spurious modes slip into the rank before the original six modes are detected. Therefore, without having prior knowledge of the system dynamics, it is extremely difficult to identify the correct sets of modes. On the other hand, Parsimonious DMD automatically identifies the six modes among these huge number of modes.

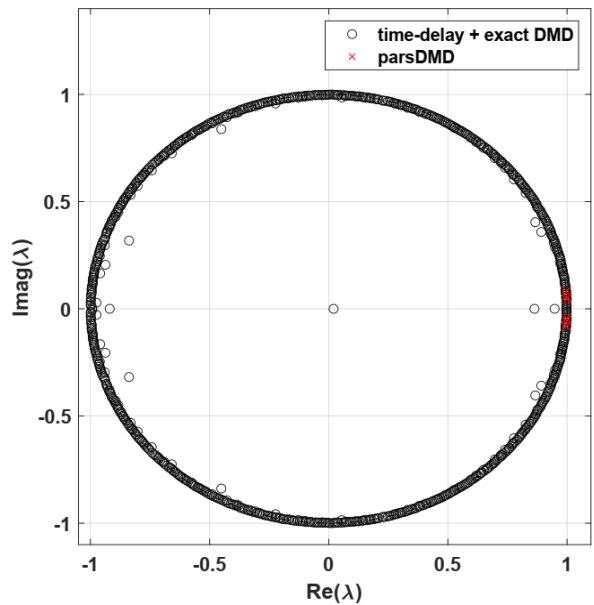

**Fig. 3**: DMD spectrum comparison of Exact DMD and Parsimonious DMD ($f_{noisy}$ with variance = 9)

The efficacy and constraints of parsimonious Dynamic Mode Decomposition (DMD) are further evaluated by introducing a substantial amount of Gaussian noise to the clean signal. Figure 4 illustrates the DMD spectrum when parsimonious DMD is applied to both the clean and the noisy signals, which exhibit varying levels of variance. A zoomed-in view of the mode conglomeration region is provided for enhanced clarity. Analysis of Figure 4 indicates that, despite the presence of significant noise, the parsimonious DMD algorithm effectively captures the true frequencies. The sole exception is one mode derived from the noisy signal with the highest variance, which is positioned slightly to the left of the unit circle, suggesting a minor decay component rather than neutral stability. Figure 5 presents the reconstruction and forecasting capabilities of the proposed algorithm when applied to the noisy signals, compared against the original clean signal, while their performance metrics are detailed in Table 1. A thorough examination reveals that the reconstruction and forecasting performance remains remarkably robust, even at a high noise variance level of 25, as demonstrated in Figure 5 and Table 1. Beyond that, parsimonious DMD is proficient in reconstructing data from $f_{noisy}$ with $\sigma^2 = 100$, while also providing a satisfactory forecasting performance. This slight performance deterioration is due to the inclusion of a damping factor because of heavy noise contamination compared to a clean signal.

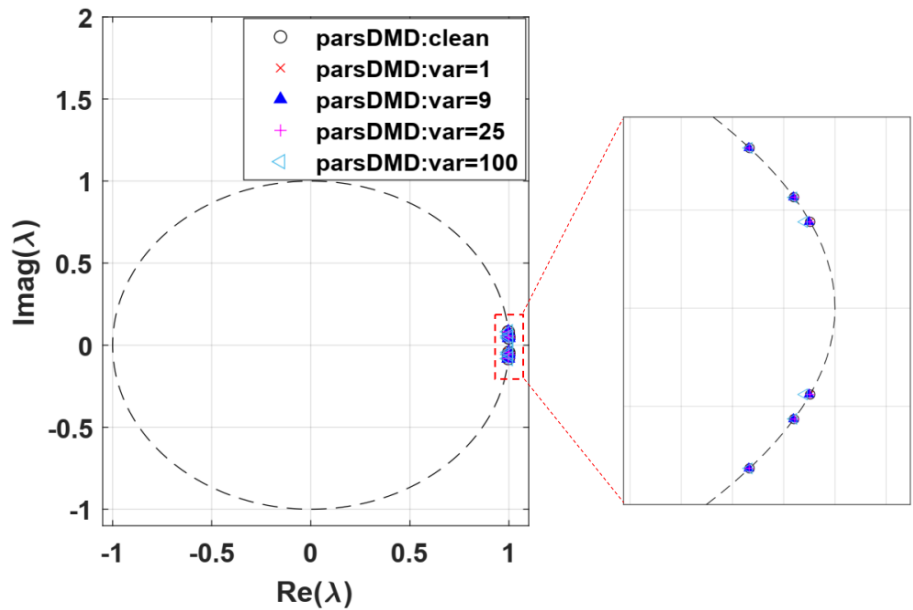

**Fig. 4**: Parsimonious DMD spectrum comparison (zoomed) of clean signal Vs Noisy signals

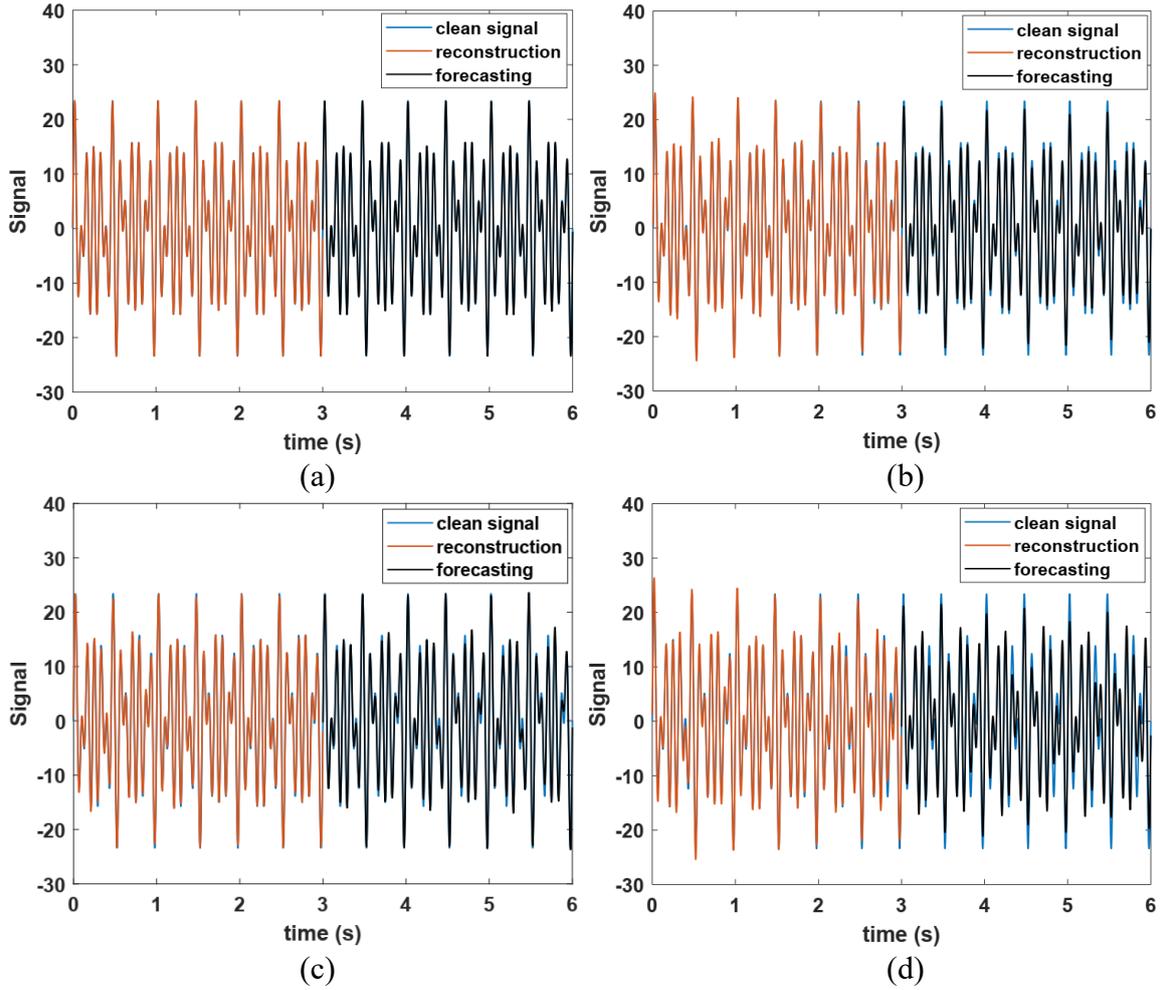

**Fig. 5**: Parsimonious DMD reconstruction and forecasting from a noisy signal with (a) var = 1 (b) var = 9 (c) var = 25 (d) var = 100

Table 1    Reconstruction and Forecasting error (synthetic time-series)

| Noise level | $\sigma^2 = 1$ | $\sigma^2 = 9$ | $\sigma^2 = 25$ | $\sigma^2 = 100$ |
|---|---|---|---|---|
| Reconstruction error (%) | 0.51 | 3.69 | 3.86 | 9.6 |
| Forecasting error (%) | 1.45 | 8.4 | 7.6 | 26 |

*4.1.2 Spatiotemporal data with hidden dynamics*

To showcase the efficacy of the developed algorithm on synthetic spatiotemporal data, a system with hidden dynamics contaminated with measurement noise is considered. This example was demonstrated by [27,60] in their paper due to the challenges associated with identifying the dynamics correctly when a rapidly decaying mode is present and the system is contaminated with measurement noise. The signal consists of two translating sinusoidal signals, where one is growing, and the other is decaying:

$$z(x,t) = \sin(k_1 x - \omega_1 t)\, e^{\gamma_1 t} + \sin(k_2 x - \omega_2 t)\, e^{\gamma_2 t} \qquad (53)$$

The values of $k_1, k_2, \omega_1, \omega_2, \gamma_1, \gamma_2$ are 1, 0.4, 1, 3.7, 1, -0.2 respectively and is kept consistent according to [27,60]. After that, high amount of measurement noise with variance $\sigma^2 = $

0.25, 1, 4, 9, 16, 25 has been added to test the algorithm. It is to be noted that the maximum amount of noise used in demonstrating the Forward-Backward DMD, Total-least-squares DMD, and Optimized DMD [27,60] was $\sigma^2 = 0.25$. Whereas in the present study, a maximum variance of 25 is implemented to test the limitation of the parsimonious DMD algorithm in identifying the hidden dynamics represented by the eigenvalue $-0.2 \pm 3.7i$. Figure 6 illustrates the efficacy of parsimonious DMD in identifying the hidden dynamics from the dataset compromised by varying levels of noise represented by their variance values. In addition, Fig. 6 addresses the dependence on the amount of data required, specifically the number of snapshots necessary for the successful identification of the decaying eigenvalue.

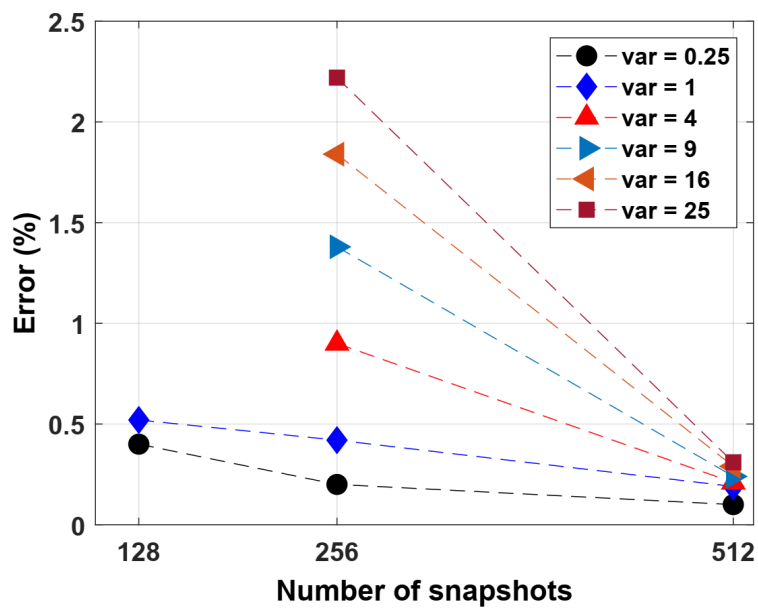

**Fig. 6**: $\ell^2$ error compared with number of snapshots for different noise variances

It is apparent from Fig. 6 that with $\sigma^2 = 0.25, 1$, just 128 snapshots are sufficient to capture the hidden eigenvalue with high accuracy (within an $\ell^2$ error margin of 0.5% for $\sigma^2 = 1$ case), and with increasing number of snapshots the error drops down to 0.1%. On the contrary, with higher variance ($\sigma^2 \geq 4$) with 128 snapshots, the algorithm is unable to detect the hidden eigenvalue, although 256 snapshots were sufficient in successful detection of the hidden eigenvalue with varying accuracy ranging from 0.9% to 2.22%. Furthermore, utilization of 512 snapshots achieved remarkable accuracy for all levels of variance values tested, and the error lies between 0.1% to 0.3%.

*4.2 Fluid Dynamics*

To illustrate the application of the parsimonious Dynamic Mode Decomposition (DMD) algorithm in fluid dynamics, two specific problems are analyzed. The first problem pertains to the classical scenario of low Reynolds number flow around a cylinder (Re=100), and the second focuses on the transonic shock buffet phenomenon that arises over a supercritical airfoil.

*4.2.1 Flow past cylinder*

2D incompressible Navier-Stokes equation is solved for flow past cylinder at $Re = 100$. Vorticity fields are considered as measurement for snapshots. Implementing the DMD algorithm identifies 5 modes which includes 1 static mode, and two mode pairs. The DMD mode shapes identified from clean data are visualized in Fig. 7.

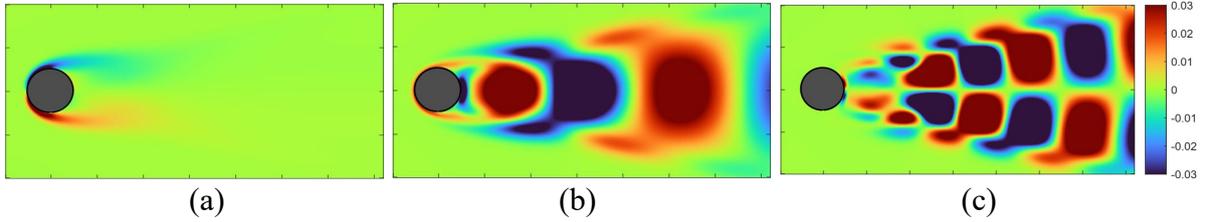

(a)  (b)  (c)

**Fig. 7**: DMD mode shapes: (a) static mode (b) mode pair 2-3 (c) mode pair 4-5

Three different types of noise have been added to the clean dataset and faithful reconstruction has been attempted using the parsimonious DMD algorithm. The first one being gaussian noise, where the clean data is contaminated with zero mean, and a variance ($\sigma^2$) of 0.1, and 0.25. The second type of noise that is added to the clean data is multiplicative noise or so-called speckle noise, where a uniformly distributed random noise with zero mean and a variance of 0.1, and 0.25 is multiplied and added to the clean dataset. Speckle noise is a significant issue in many experimental imaging techniques, including PIV, ultrasound, radar, and optical coherence tomography. This noise can obscure important details and complicate data analysis, making effective noise reduction techniques crucial in obtaining accurate and meaningful results from experimental data. The mathematical formulation for speckle noise:

$$X_{noisy} = X_{clean} + \mathcal{N}(0, \sigma^2) * X_{clean} \qquad (54)$$

and finally, a salt and pepper noise, also called impulse noise with 2%, and 5% pixel density corruption has been added, i.e. for 2% pixel corruption, the number of pixels corrupted $I = 0.02 * numel(X_{clean})$, where $numel$ function represents the number of element in the dataset $X_{clean}$. This type of noise is a common occurrence in experimental data, particularly in imaging techniques such as Particle Image Velocimetry (PIV) [13,61]. It arises from various sources, including sensor defects, transmission errors, and environmental factors, and can significantly affect the accuracy of the measured velocity fields. At first, $I$ number of pixels are chosen randomly based on the density corruption criterion. After that, the first half of those chosen pixels are denoted a value of zero, and the rest of the chosen pixels are denoted a value based on the maximum value in the dataset. Therefore, $X_{noisy}$ is a combination of $X_{clean-dn}$ and $X_{dn}$, such that,

$$X_{dn} = \begin{cases} 0, & 1 \leq dn \leq \lfloor I/2 \rfloor \\ \max(X_{clean}), & \lfloor I/2 \rfloor + 1 \leq dn \leq I \end{cases} \tag{56}$$

where $\lfloor . \rfloor$ is the floor function, and $X_{clean-dn}$ are pixels from the clean dataset that has not been corrupted.

Figure 8 illustrates a vorticity field snapshot for clean data, and the effect due to the contamination of Gaussian, multiplicative, and salt and pepper noise, and finally the same snapshot after implementation of parsDMD reconstruction. In all of the cases, the parsDMD algorithm automatically identifies 5 modes just as it identified from clean dataset. Moreover, the reconstruction from Gaussian, and multiplicative noise does an excellent job, the reconstructed snapshot from salt and pepper-contaminated noise is still quite noisy. Nevertheless, the effect of noise contamination is reduced which is clearly visible. It is to be noted that the most effective way to deal with salt and pepper-type noise or impulse noise is to preprocess the data with RPCA algorithm [62]. Although computationally expensive, it was shown that filtering the salt and pepper-contaminated data with RPCA, and subsequently implementing DMD provides excellent results [13,61,63].

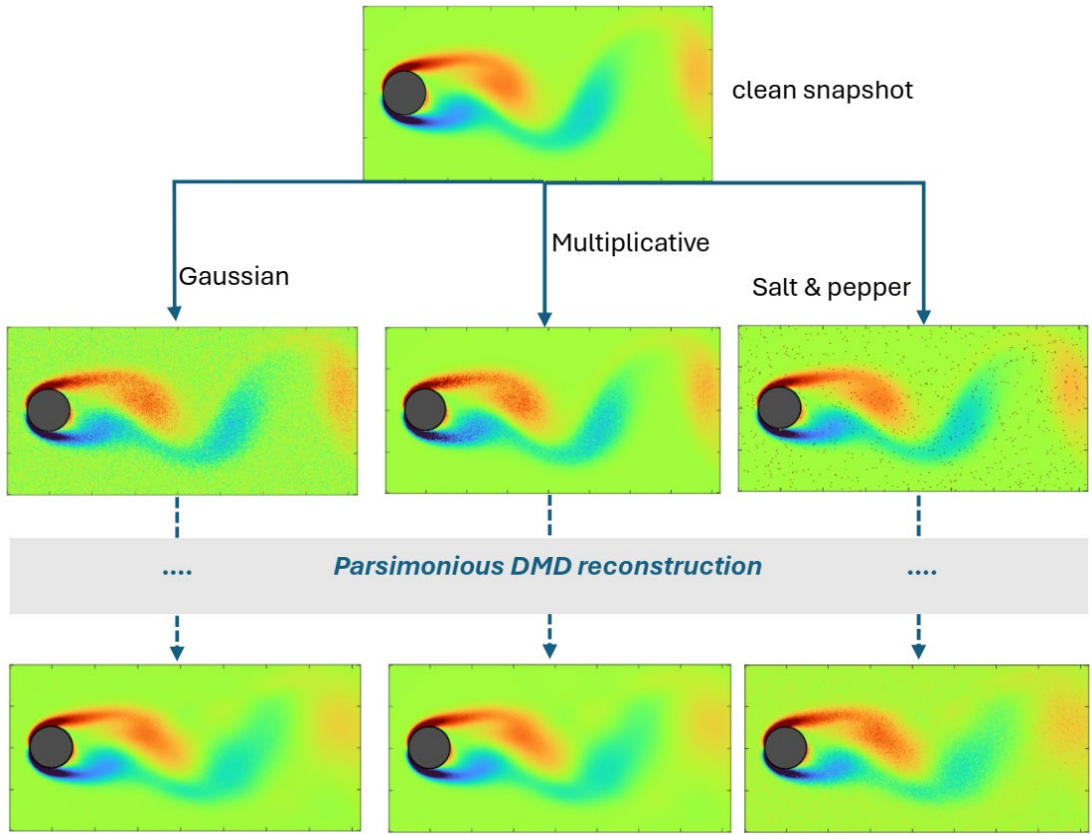

**Fig. 8**: Noise-contaminated snapshots and their parsimonious DMD reconstructions

The discrete-time DMD eigenvalues are depicted in the DMD spectrum shown in Fig. 9. To effectively illustrate the impact of noise contamination, a zoomed-in view of the areas surrounding mode 2 and mode 4 has been provided. It is evident that the frequencies of the modes are accurately represented despite the presence of noise. However, in instances of salt and pepper noise, the modes are observed to be slightly displaced from the unit circle, reflecting their decaying characteristics. Ideally, these modes should lie on the unit circle, indicating neutral stability; thus, preprocessing of the contaminated dataset using the RPCA algorithm, as previously discussed, is necessary. Nonetheless, the ability of parsDMD to manage datasets affected by various complex noise types is noteworthy.

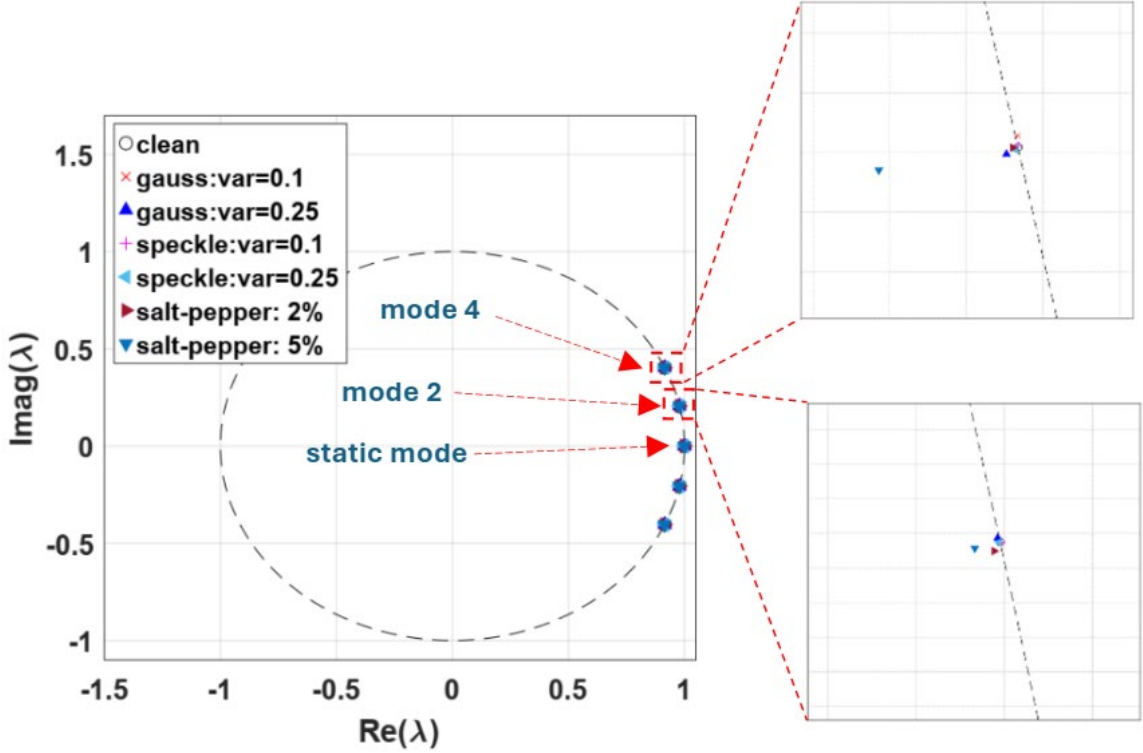

**Fig. 9**: DMD spectrum for clean flow-field data compared with noise contaminated data

*4.2.2 Transonic buffet*

Within the very narrow regime of transonic flow, complex interactions between shock waves and boundary layer cause some airfoil/wing configurations to experience self-sustained shock oscillations that induce large aerodynamic load variations. Such oscillations, referred to as transonic shock buffet, can cause a reduction in stability, control, handling qualities, and structural integrity, thereby limiting the aircraft buffet envelope. Due to the complexity and the severity of this phenomenon, it has been an active topic of research for the last few decades, which requires either expensive experimental investigations in the wind tunnel or high-fidelity computations, which, due to the inherent large-scale nature of the models, demands extensive computational resources. In this work, numerical investigation for analyzing transonic buffet has been conducted on OAT15A supercritical airfoil.

Measurements from the experimental wind tunnel tests conducted by ONERA are available [64] and are used for the CFD simulation. Governing equations were discretized and solved based on commercially available finite volume method-based CFD software ANSYS Fluent 19 [65]. SST $k - \omega$ has been chosen as the turbulence closure model. A non-dimensional time-step of $\Delta \tau = 0.004$ has been selected for further computations where $\Delta t, U_\infty, c$ are time step, freestream speed, and chord length of airfoil respectively, where $(\Delta \tau = \Delta t U_\infty / c)$. For more details about grid selection, spatial and temporal convergence studies, and validation from experimental counterpart, the readers are directed towards [13,15,66]. For showcasing in the current study, the benchmark test case at operating conditions of $M_\infty = 0.73$ and $\alpha = 3.5^o$, at speed-of-sound of $a_\infty = 330\ m/s$ has been selected. Flow-field of Mach number contours are chosen for the DMD analyses. The initial five buffet cycles from the dataset are utilized for training purposes, while a comprehensive time-series dataset comprising 40 buffet cycles is selected to evaluate the forecasting capabilities of the parsDMD algorithm. Additionally, a comparative analysis with Exact DMD is included.

Parsimonious DMD identifies 5 modes, among them are 1 static mode representing mean flow-field characteristics, mode pair 2-3 is the buffet mode, i.e. the frequency of mode 2-3 is the same as the shock oscillation frequency, and mode pair 4-5 captures the first superharmonic of the buffet frequency. Modes beyond the fifth, mode pair 6-7 and higher, represent the second, third, fourth, and additional superharmonics, having negligible contribution [66] towards the flow-field, are systematically truncated by the parsimonious DMD mode selection criterion. The mode shapes are shown in Fig. 10, mode pair 2-3 has two coherent structures: one is in the shock oscillatory region that covers the total shock travel distance and represents the movement of the shock to its minimum and maximum excursion, while the other emanates from the shock foot and extends to trailing edge and beyond. Modes 4-5 represent two clear, coherent structures at the shock oscillatory region of opposite strength and propagate upward into the shock and then to the far-field Another comparatively smaller structure which originates at the shock foot at its maximum position and propagates along the upper surface of the airfoil into the wake region and represents the interaction of boundary layer and trailing edge vortex shedding. For more details into the shock buffet characteristics and the temporal evolution of DMD modes, the readers are directed towards [15,66].

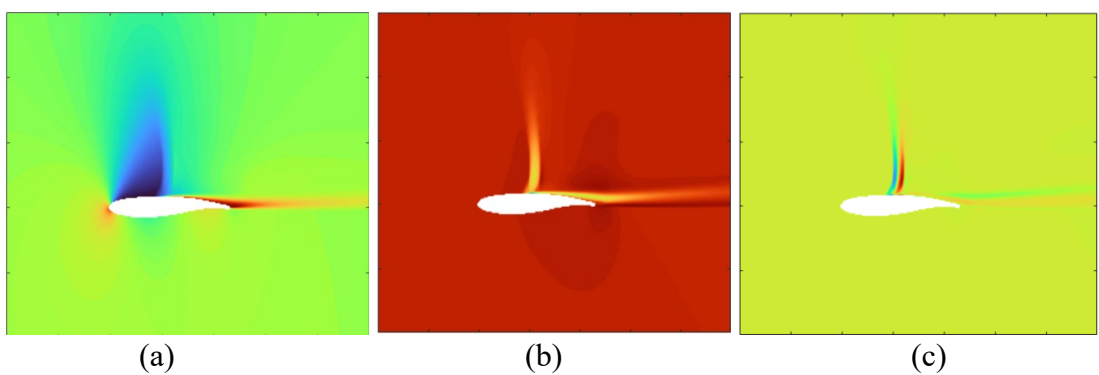

(a)          (b)          (c)

**Fig. 10**: DMD mode shapes of transonic buffet: (a) static mode (b) mode pair 2-3 (c) mode-pair 4-5

The reconstruction and forecasting ability of parsDMD (5 modes) and its comparison with Exact DMD with 21 modes was evaluated. The Frobenius error for reconstruction for parsDMD is 1.46% compared to 0.89% in the case of Exact DMD, while for forecasting, the values are 1.5% and 1.46% respectively. It is quite striking that while in the case of parsDMD the error jumped from 1.46% to 1.5%, for Exact DMD its 0.89% to 1.46%. Therefore, to delve deeper, the dynamical error, i.e. the evaluation of rms error per snapshot basis for reconstruction as well as forecasting, is illustrated in Fig. 11. The dynamical error is evaluated at a particular time-step $j$ is defined as,

$$Err_{dyn_j} = \|x_j - x_{recon_j}\|_2 / \|x_j\|_2 \qquad (57)$$

where $x_j$ is the jth snapshot from CFD data, and $x_{recon_j}$ is the jth snapshot from DMD reconstruction/forecasting. A clear trend is evident from the figure is that parsDMD is much more stable compared to Exact DMD. Although initially, during the reconstruction phase, Exact DMD has lower error compared to parsDMD as denoted by the zoomed-in portion of the reconstruction zone, its zoomed-in counterpart for the forecasting region identifies exactly the opposite phenomenon where the parsDMD outperforms the Exact DMD algorithm.

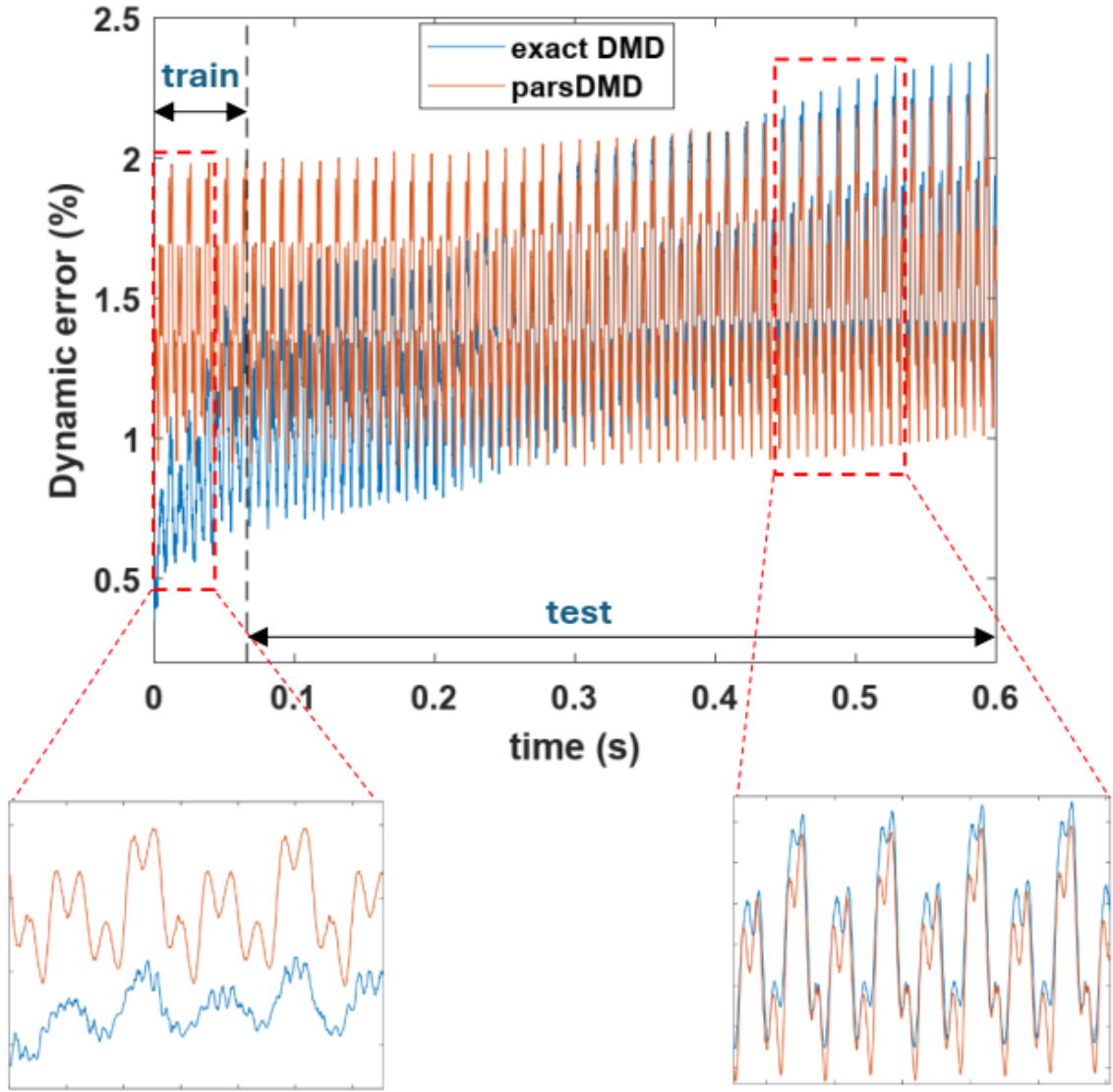

**Fig. 11**: Dynamic error comparison for parsDMD and Exact DMD

This difference in stability is much more prominent when shock buffet data is contaminated with noise. Even with the slightest inclusion of noise (gaussian with $\sigma^2 = 0.1$), the eigenvalues identified by exact DMD shows large bias (Fig. 12a), while the parsDMD identified eigenvalues are true to clean data. For the sake of brevity, only the zoomed-in portion of the DMD spectrum where modes conglomerate is shown. Furthermore, observing the dynamic error evolution reveals that the reconstruction, and eventually the forecasting error increases exponentially for Exact DMD in comparison with parsDMD, in which case the dynamic error is far more stable, as shown in Fig. 12b. This is primarily due to the ease of inclusion of time-delay embedding which is augmented in the parsDMD algorithm that helps reducing the bias in eigenvalues, and the true eigenvalues of the system are detected, and the spurious ones are truncated by the parsDMD mode selection criterion.

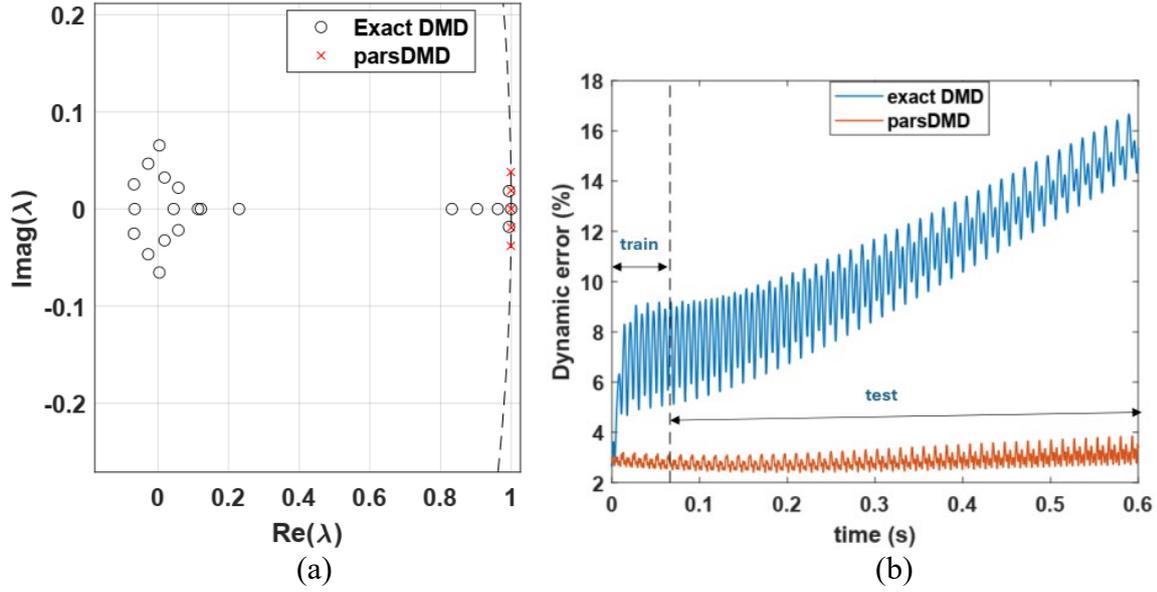

**Fig. 12**: (a) zoomed-in DMD spectrum comparison for noisy buffet data (Exact Vs parsDMD) (b) Dynamic error performance (Exact Vs parsDMD)

*4.3 Comparison with sparsity-promoting DMD*

In order to understand the difference between the proposed parsimonious DMD algorithm showcased in this paper and the sparsity-promoting DMD [48] (spDMD) algorithm, it is imperative to understand how spDMD algorithm works. The key difference is how Eq. (31) is being solved. In spDMD a sparsity structure is invoked in the cost function (Eq. (31) such that a subset of DMD modes can be identified using the ADMM (Alternating Direction Method of Multipliers) algorithm [67], the equation is rewritten for convenience,

$$\underset{b}{\textbf{minimize}} J(\textbf{b}) = \|X - \Phi D_b V_\lambda\|_F^2 + \gamma \|b\|_1 \tag{58}$$

A set of user-defined sparsity values ($\gamma$) needed to be incorporated into the cost function equation. The best way to do that is to select the lowest sparsity value that puts weightage on all of the modes (just like while solving for standard mode selection procedure using the least-squares method), and the extreme end of sparsity value, which puts weightage only on one mode, and the rest of the amplitudes are zero. Thereafter, several sparsity values (generally 200) are selected which are logarithmically spaced between these two extreme values. The following bar charts in Fig. 13 show the effect of increasing sparsity values on amplitudes showcased in transonic buffet case. It is identified that with the lowest sparsity ($\gamma = 0.1$) value, all of the modal amplitude has some non-zero values, but as the $\gamma$ value is increased, the sparser the structure becomes and with the highest sparsity value provided (shown in Fig. 13d) only one mode retains non-zero amplitude and the rest are of zero amplitude.

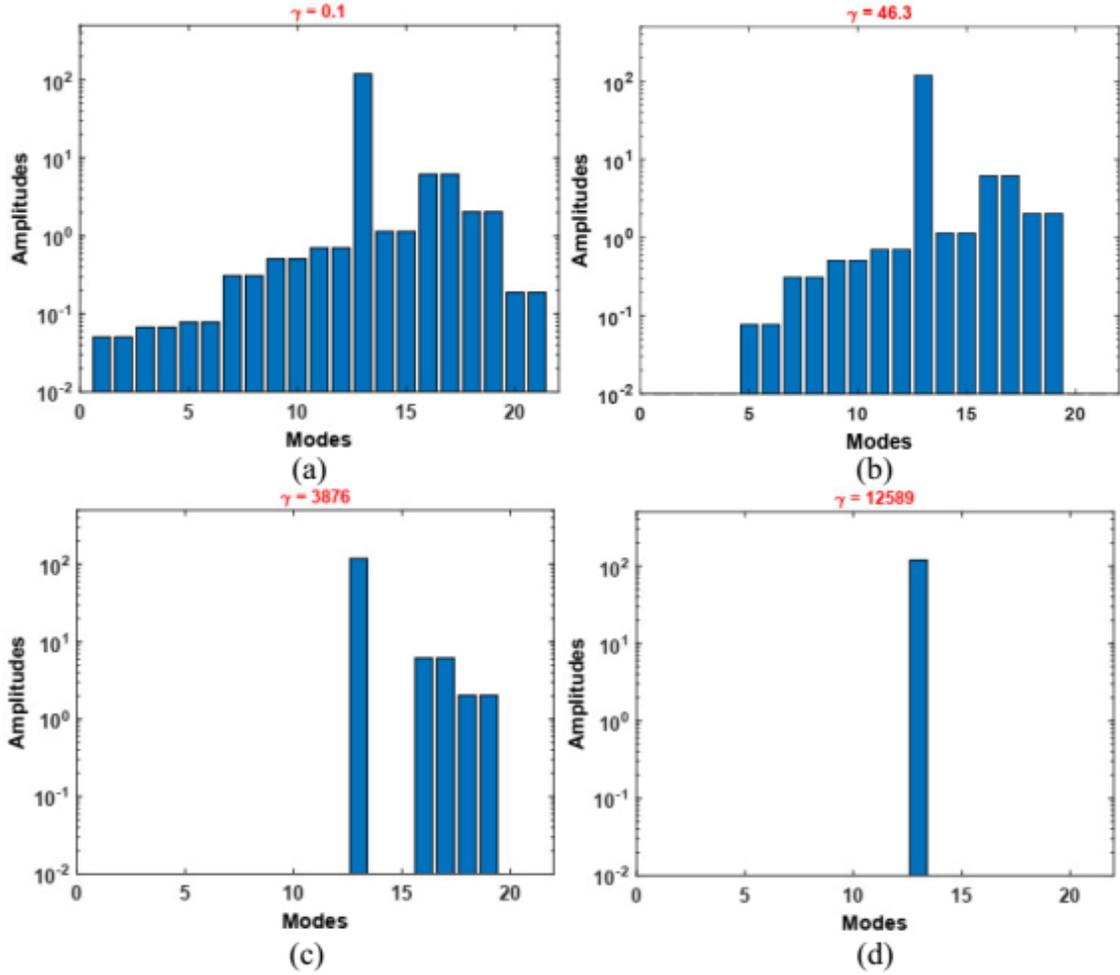

**Fig. 13**: Effect of varying sparsity values of mode amplitudes: (a) $\gamma = 0.1$ (b) $\gamma = 46.3$ (c) $\gamma = 3876$ (d) $\gamma = 12589$

To identify the optimal sparsity value that leads to the optimal sparse subset of modes, two things need to be considered: the variation of the number of non-zero modal amplitudes w.r.t. $\gamma$, and the variation of performance loss w.r.t $\gamma$. The performance loss in percentage is defined as $P_{loss} = 100 * \sqrt{\frac{J(b_{sp})}{J(b)}}$ and the cost function $J$ is defined as $J(\mathbf{b}) = \mathbf{b}^*\mathbf{P}\mathbf{b} - \mathbf{q}^*\mathbf{b} - \mathbf{b}^*\mathbf{q} + s$. As expected, increasing $\gamma$ values leads to a lesser number of modes, but on the other hand, performance loss goes up as the model becomes sparser. To select the optimal sparse subset of amplitudes and construct a ROM it is necessary to preserve the parsimony of the model. Therefore, it is useful to compare the trends between the number of non-zero amplitudes and performance loss w.r.t $\gamma$ (Fig. 14) and identify the intersection point of the patterns that indicate the best possible sparsity which leads to the optimally sparse set of modal amplitudes. In this particular case, it leads to a $\gamma$ value of 3867.6. According to the identified $\gamma$ value from Fig. 14, only 5 modes have non-zero amplitudes (the same result was obtained previously with parsDMD).

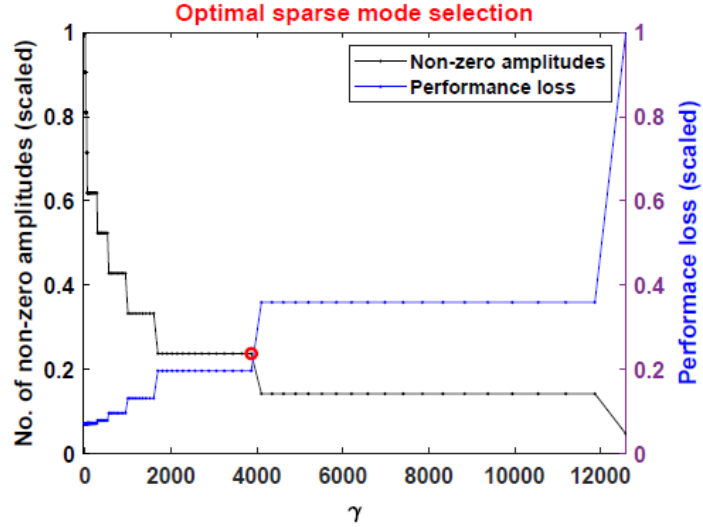

**Fig. 14**: Optimal sparser mode selection based on sparsity structure and performance loss

Although an effective algorithm in identifying the sparse DMD modes, the main limitation for spDMD algorithm is identifying the extremum sparsity ($\gamma$) values, i.e. $\gamma$ value for all modes (in the showcased example 0.1), and $\gamma$ for 1 mode solution, which was 12589. These values are evaluated by hit-and-trial basis, and it is imperative that these extremum sparsity values are correctly evaluated for the accuracy of the algorithm in identifying the right number of modes. Moreover, the sparsity values change drastically from one system to another, even for the same dynamical system when the POD rank truncation is different, or the data is contaminated with noise etc. Therefore, identifying the extremum sparsity values is a tedious but essential process, which makes the process not automatic. Furthermore, after evaluating the extremum sparsity values, the spDMD algorithm uses a set of sparsity values logarithmically spaced between these two extrema values and utilizes the ADMM algorithm to solve them. ADMM algorithm gets progressively computationally expensive as the sparsity value goes up, i.e. for higher $\gamma$ values that identifies lesser number of modes takes more time to solve the equation with low $\gamma$ values. On the other hand, in the case of parsimonious DMD algorithm, where the sparse identification of DMD modes is obtained by utilizing the OMP algorithm, is significantly faster. This is primarily since OMP identifies the sparsest solution at first (1 mode solution), and then progressively moves towards a higher number of modes. In addition, the implementation of scaled sparsity parameter and the convergence criterion described in Eq. (42) and (43) ensures the algorithm stops when convergence is achieved. Therefore, this algorithm bypasses the requirement of any pre-determined estimation of sparsity values, making it robust, totally automatic and optimally sparse in nature.

The implementation of spDMD has been demonstrated through several examples previously discussed, highlighting its limitations in comparison to parsDMD. As illustrated in Table 2, the range of sparsity values ($\gamma_{min}, \gamma_{max}$) exhibits significant variability across different dynamical systems. Furthermore, even within the same system, minor additions of noise lead to substantial discrepancies in these values. The computational time recorded for spDMD in Table 2 reflects the duration required after determining $\gamma_{min}$ and $\gamma_{max}$ through a laborious trial-and-error approach. A grid search technique could be employed to ascertain the sparsity values. However, it is important to recognize that the considerable differences in scale would render this process more time-consuming. In contrast, parsDMD does not necessitate such procedures, resulting in a markedly reduced total computational time when compared to spDMD. For instance, in the case of flow past a cylinder, parsDMD operates five times faster than spDMD, even after the sparsity value has been identified. The identification of sparsity values for the cylinder scenario takes approximately one minute, indicating that, in practice, parsDMD is over 200 times faster than spDMD. Additionally, the performance of sparsity-promoting DMD in accurately identifying modes deteriorates with the introduction of noise. In all tested scenarios, spDMD tends to overestimate the number of required modes when data is noisy, whereas parsDMD successfully identifies the modes in comparison to a clean dataset, thereby demonstrating greater robustness.

Table 2: Comparison between spDMD and parsDMD

| Method | Sparsity-promoting DMD | | | parsDMD |
|---|---|---|---|---|
| Test cases | $\gamma_{min}$ | $\gamma_{max}$ | Computational Time after estimating the sparsity values (s) | Total Computational time (s) |
| Cylinder | 0.0316 | 50119 | 1.5 | 0.3 |
| Cylinder + $\mathcal{N}(\sigma^2 = 0.25)$ | 31.62 | 100000 | 1.5 | 0.3 |
| Buffet | 0.1 | 12589 | 8.6 | 3.7 |
| Buffet + $\mathcal{N}(\sigma^2 = 0.1)$ | 0.4 | 79433 | 13 | 3.8 |

*4.4 Atmospheric Science (Sea-Surface Temperature)*

In atmospheric science research, monitoring the sea-surface temperature (SST) is critically important because of its key role in regulating Earth's climate and influencing various atmospheric processes including driving atmospheric circulation, and formation of weather systems. Moreover, SST is a key metric in climate change indicators. Monitoring

SST helps scientists understand the effects of climate change, predict extreme weather events, and model atmospheric processes that impact both regional and global climates.

In this study, NOAA optimum interpolation SST data [68] is used for parsDMD analysis. The dataset contains latitude, longitude, and monthly mean of sea-surface temperature at those locations spanning from the year 1980-2023. The first 400 snapshots (roughly 33.3 years) are used for training the DMD algorithm, and the rest of the 116 snapshots (roughly 9.7 years) are kept to assess the forecasting accuracy of the trained model. Implementation of parsDMD on the training dataset captured 5 modes: a static mode and 2 dynamic mode pairs (illustrated in Fig. 15). It should be noted that the dynamic mode pairs have a certain positive growth rate associated with them, indicating the rise in global SST. It was observed that with just 5 modes, the Frobenius error for reconstruction and forecasting is barely 3.53% and 4.1% respectively.

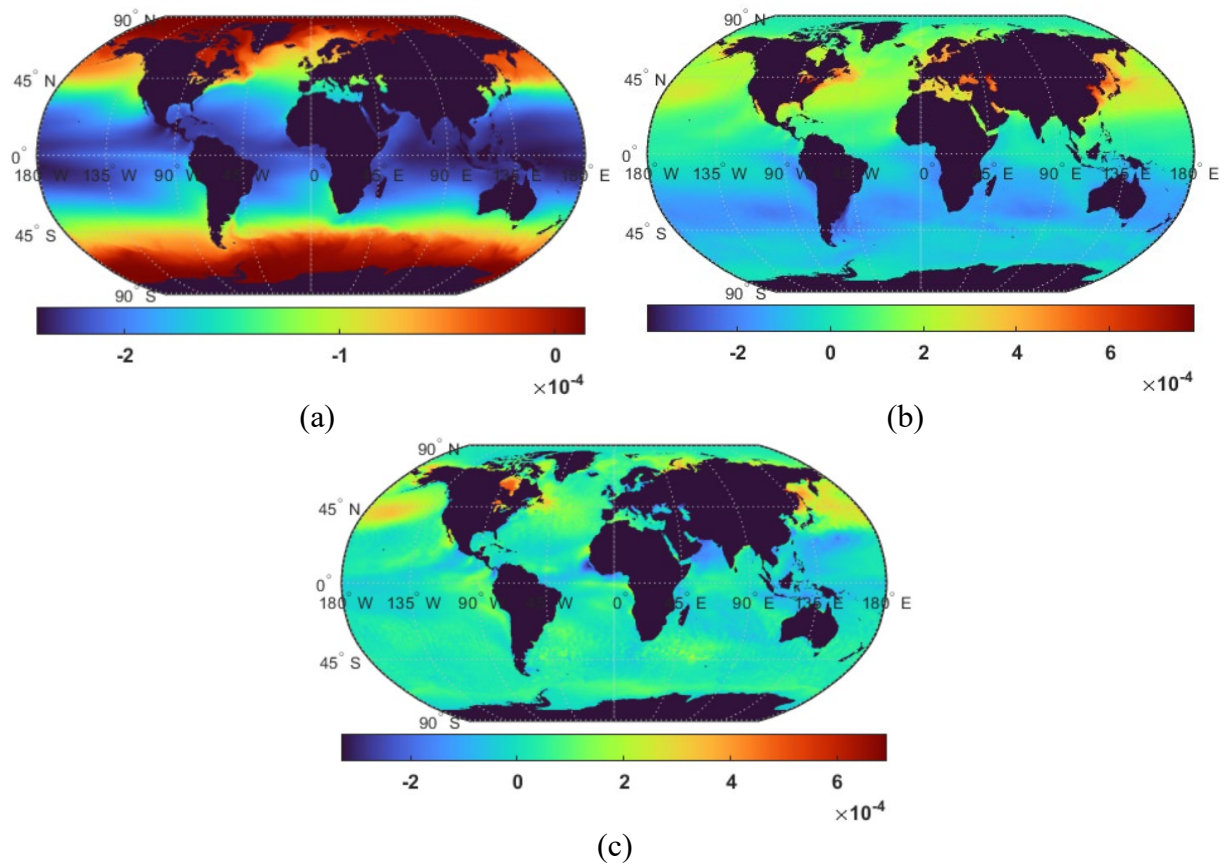

**Fig. 15**: DMD mode shapes (SST): (a) static mode (b) mode pair 2-3 (c) mode pair 4-5

An in-depth evaluation of the forecasting capability of the trained model is conducted by analyzing the dynamic error, as depicted in Fig. 16a. In addition, a side-by-side comparison is made between the final SST snapshot from the observed data and the last snapshot produced by the parsDMD prediction. The trend in error clearly demonstrates that the dynamic error is confined within the range of 3.2% to 5.5%, which is deemed satisfactory. Furthermore, the disparity between the observed and predicted SST snapshots is nearly imperceptible, implying that parsDMD could serve as an effective instrument for the analysis of SST data.

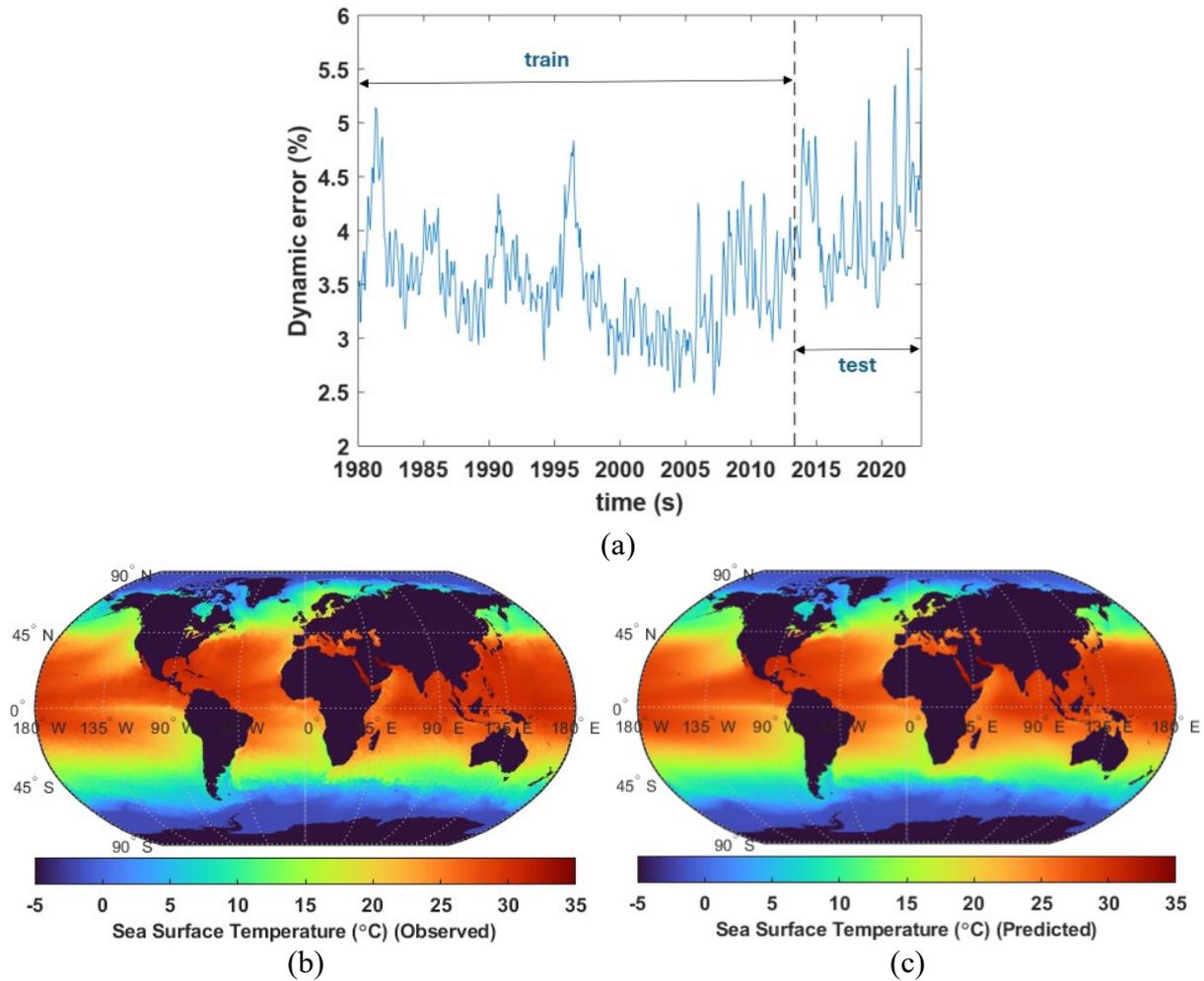

**Fig. 16**: (a) Dynamic error for SST reconstruction and forecasting (b) Observed SST snapshot
(c) Predicted SST snapshot

## 5. Conclusions

This work introduces a novel Parsimonious Dynamic Mode Decomposition (parsDMD) algorithm designed to automatically select an optimally sparse subset of dynamic modes from complex dynamical systems datasets. The algorithm integrates time-delay embedding to handle noise and nonlinearity, and uses Orthogonal Matching Pursuit (OMP) with a custom scaling criterion to automatically identify the most relevant modes, ensuring efficient and accurate reconstruction of system dynamics. By eliminating the need for parameter tuning to determine sparsity values, parsDMD simplifies the mode selection process, offering an intuitive and computationally efficient approach to reduced-order modeling. Its automated nature makes it particularly well-suited for real-time diagnostics and forecasting in various complex systems, enhancing both speed and reliability across different applications.

The effectiveness and versatility of parsDMD have been demonstrated across multiple datasets, including synthetic spatiotemporal and purely temporal data, fluid dynamics simulations (e.g., flow past a cylinder and transonic buffet), and real-world atmospheric data. Its application to the sea surface temperature (SST) dataset from atmospheric science underscores the algorithm's ability to handle large-scale, real-world systems. In this context, parsDMD successfully captures the key modes from SST data, offering a computationally efficient framework for studying long-term patterns in ocean temperatures. The diverse applications, ranging from synthetic data to SST analysis, further highlight parsDMD's ability to deal with different types of complex dynamical systems, as the algorithm consistently outperforms existing methods in accurately identifying dominant modes while maintaining a high degree of forecasting capability.

Additionally, parsDMD's resilience to noise (including Gaussian, multiplicative, and salt-and-pepper noise) has been validated across different datasets, demonstrating its robustness even in challenging environments. This robustness makes parsDMD a powerful tool for real-time diagnostics, reduced-order modeling, and forecasting in applications where noise contamination is a common challenge, such as in atmospheric science, climate research, and fluid dynamics. By ensuring accurate mode identification and providing strong performance even in noisy data environments, parsDMD extends its applicability beyond traditional DMD approaches.

In summary, parsDMD represents a significant advancement in the field of data-driven dynamical system modeling using DMD-based methodologies. By removing the need for manual sparsity tuning and delivering superior performance across both spatiotemporal and purely temporal datasets, it addresses the limitations of previous methods for the identification of sparse subsets of DMD modes. Its successful application to complex systems like transonic buffet, and SST data in atmospheric science, where accurate modeling of climate variability is crucial, further validates its practical utility. The algorithm's automated nature, combined with its ability to generalize to a range of applications, makes it an ideal tool for real-time diagnostics, reduced-order modeling, and forecasting across multiple disciplines, enhancing both the speed and accuracy of data-driven models.

**Acknowledgments**

The authors would like to thank the Defence Science and Technology Group of the Australian Department of Defence for partially supporting this research and Dr. Matteo Giacobello for insightful discussions.